\documentclass[nohyper,12pt,letterpaper]{JHEP}

\def\be{\begin{equation}}
\def\ee{\end{equation}}
\def\bear{\begin{eqnarray}}
\def\eear{\end{eqnarray}}
\def\nn{\nonumber}

\newcommand\px[1]{{\partial_{#1}}}
\newcommand\qx[1]{{\partial^{#1}}}

\def\BZ{{{\bf Z}}}
\def\BC{{{\bf C}}}
\def\BR{{{\bf R}}}
\newcommand\MR[1]{{{\bf R}^{#1}}}               
\newcommand\MS[1]{{{\bf S}^{#1}}}               

\newcommand\SUSY[1]{{{\cal N}= {#1}}}           

\def\u{{\mu}}
\def\v{{\nu}}

\def\th{{\theta}}
\def\Th{{\Theta}}

\title{M(atrix)-Theory Scattering\\
 in the Noncommutative $(2,0)$ Theory}
\author{Ori J. Ganor and Joanna L. Karczmarek\\
Department of Physics, Jadwin Hall \\
Princeton University \\
NJ 08544, USA}

\abstract{
We study the Quantum-Mechanics on the hyper-K\"ahler manifold
that is the blow-up of an $A_1$-singularity.
This system is relevant for M(atrix)-theory as it was conjectured
to describe scattering in the 
``noncommutative'' deformation of a free 5+1D tensor multiplet
in the sector with two units of longitudinal light-like momentum.
}

\received{???????? ?st, 2000} \accepted{???????? ?th, 2000}

\preprint{\hepth{0007166}; PUPT-1943}
\begin{document}
             

\section{Introduction}\label{intro}
A single M5-brane is described at low-energies by a free tensor
multiplet with $\SUSY{(2,0)}$ supersymmetry.
An extension
of the free tensor multiplet to an interacting theory has been 
suggested in \cite{ABS} and \cite{NekSch,Berk}.
It was motivated by the string-theory realization \cite{CDS,DH}
of gauge theories on a noncommutative $\MR{4}$.
This conjectured 5+1D theory breaks Lorenz invariance explicitly.
It is assumed to depend
on a constant anti-self-dual 3-form parameter $\Th^{ijk}$ of dimension
${\mbox Mass}^{-3}$.
At low-energies the theory reduces to the free tensor multiplet.

In \cite{SWNCG} a consistent limit of the M5-brane with a strong 3-form
field-strength was presented and it was suggested that at low-energies
this limit describes a decoupled ``noncommutative'' $(2,0)$-theory.

Recently, the theory has been re-incarnated in 
OM-theory \cite{GMSS}.
 There, a limit with a critical time-like component, 
$H_{012}$,  was studied and was argued to yield
a decoupled theory. At low-energies it reduces to a free
tensor multiplet.
Presumably, this is the ``noncommutative'' $(2,0)$ theory with
nonzero $\Th^{345} = \Th^{012}$.
In \cite{GMSS}, time-like noncommutativity,
critical electric fields and space-like noncommutativity are
related in a unified structure.
(See also \cite{HVerl}-\cite{AhGoMe} for related results.)
The theory of \cite{ABS,NekSch,Berk} is a decoupled sector
of a special case of
OM-theory, for which $\Theta^{ijk}$ is light-like (see \cite{AhGoMe}
for a nice discussion).

The action of a $U(1)$ gauge-theory on a noncommutative $\MR{4}$,
can be expanded as the free action plus
terms that are of higher order (in the noncommutativity) and
depend only on the field-strength $F_{ij}$ and its derivatives
\cite{SWNCG}.
Similarly, the equations of motion of the noncommutative 5+1D theory
can, presumably, be expanded in $\Th^{ijk}$.
We will derive the leading terms in section (\ref{motiv}).

One of the exciting features about
this interacting theory is that it has
a conjectured M(atrix)-model \cite{BFSS} that is described in terms of 
a quantum mechanics on a nonsingular space \cite{ABS,NekSch,Berk}.

The simplest nontrivial sector of this M(atrix)-model is
the sector with longitudinal, light-cone, momentum of $p_{\|}=2/R_{\|}$,
where $R_{\|}$ is the radius of the light-like direction.
It describes the scattering of two massless
particles, corresponding to the tensor multiplet,
each with longitudinal, light-cone, momentum of $p_{\|}=1/R_{\|}$.
This is similar to the calculations of scattering of gravitons
and their supersymmetric partners \cite{DKPS,PSS,RamTay}
in the M(atrix)-model of 10+1D M-theory.
The M(atrix)-model for the noncommutative
tensor multiplet with $p_{\|}=2/R_{\|}$ is described by
quantum-mechanics on a blown-up $A_1$-singularity.
In this paper we will study this quantum-mechanics and calculate
the low-energy scattering, in the quantum-mechanics.

The paper is organized as follows.
In section (\ref{motiv}) we describe the 
leading order low-energy limit of the noncommutative
M5-brane theory. That is our motivation for studying the
quantum-mechanics on a blown-up $A_1$ singularity.
In section  (\ref{qmaone}) we describe in details the
Quantum Mechanics on the blown-up $A_1$ singularity.
In subsection (\ref{scat}) we calculate the s-wave scattering
amplitude on the $A_1$ singularity.
In the appendix, we present the calculation for scattering of
two scalar particles in field-theory, up to order $O(\Theta)^2$.


\section{Motivation: noncommutative M5-brane}\label{motiv}
One motivation for studying the QM on the blown-up $A_1$-singularity
is that it is the M(atrix)-model for the noncommutative deformation
of a free 5+1D tensor-multiplet.
We will now describe this theory at lowest order in the ``noncommutativity.''

\subsection{Free M5-brane}
A single M5-brane is described, at low-energies, by a free tensor
multiplet of $\SUSY{(2,0)}$ supersymmetry.
The bosonic fields are 5 free scalars, $\phi^I$ ($I=1\dots 5$) and
a 3-form tensor field-strength $H_{ijk}$ with equations of motion:
$$
H_{ijk} = -\frac{1}{6}\epsilon_{ijklmn}H^{lmn},\qquad
\px{\lbrack l}H_{ijk\rbrack}=0.
$$
Here, indices are lowered and raised with the metric:
$$
\eta_{mn}dx^m dx^n = -dx_0^2 +dx_1^2 +dx_2^2 +dx_3^2 +dx_4^2 +dx_5^2,
$$
and the anti-symmetric $\epsilon$-symbol is normalized such that:
$\epsilon_{012345}=+1$. In particular, we have: $H_{012} = H_{345}$.
The notation $\lbrack ij\dots k\rbrack$ means complete anti-symmetrization.
Thus:
$$
T_{\lbrack i_1\dots i_n\rbrack} \equiv \frac{1}{n!}\sum_{\sigma\in S_n}
(-)^\sigma T_{i_{\sigma(1)}\dots i_{\sigma(n)}}.
$$
The normalization is such that on 3-cycles:
$$
\int H_{ijk} dx_i dx_j dx_k \in 2\pi \BZ.
$$

Later on, we will need the propagator:
\be\label{propHH}
\langle H_{ijk}(x) H_{lmn}(y)\rangle =
\int {{d^6 p}\over {(2\pi)^6}} e^{i p (x-y)} G_{ijk;lmn}(p).
\ee
We calculate it by adding a self-dual part to $H_{ijk}$ 
and writing it as $H_{ijk} =3\px{\lbrack i} B_{jk\rbrack}$ with the action
(See for instance \cite{GanMot} for details):
$$
-\frac{1}{24\pi}\int H_{ijk} H^{ijk} d^6 x.
$$
We then keep only the anti-self-dual part of the propagator.
The result is:
\bear
G_{ijk;lmn}(p) &=&
{{18\pi}\over {p^2 + i\epsilon}}
  \eta_{rl}\eta_{sm}\eta_{tn}
  \left(p_{\lbrack i} p^{\lbrack r}
    \delta_j^s\delta_{k\rbrack}^{t\rbrack}
  -\frac{1}{6} {\epsilon_{ijk}}^{\sigma\lbrack st} p^{r\rbrack}
\right)
\nn\\ &&
+{{\pi}\over {2}}\left(
\epsilon_{ijklmn} - 6\delta^{\lbrack r}_{\lbrack i}\delta^s_j
\delta^{t\rbrack}_{k\rbrack}\eta_{lr}\eta_{ms}\eta_{nt}
\right).
\label{Gp}
\eear

\subsection{The interacting theory}
The interacting theory is described by interactions $L_{int}(\Th)$
that depend on a constant anti-self-dual 3-form $\Th^{ijk}$.
It satisfies:
$$
\Th_{ijk} = -\frac{1}{6}\epsilon_{ijklmn}\Th^{lmn}.
$$
$L_{int}$ involves the fields $\phi^I$, $H_{ijk}$, the fermions and
their derivatives. 

To first order in $\Th$, the interactions can be described
by a self-dual dimension-9 operator, ${\cal O}_{ijk}$, in the free theory.
The interaction is $\delta L = \Th^{ijk}{\cal O}_{ijk}$.

The bosonic part of the interaction turns out to be:
\bear
\delta L &=&
  \frac{1}{96\pi}\Th^{ijk}H_{ijl} H^{mnl}H_{mnk}
  +\frac{1}{4}\Th^{ikl}H_{jkl}\px{i}\Phi^I\qx{j}\Phi^I
  + O(\Theta)^2.
\label{cubic}
\eear
Here, $H$ should not be confused with the critical asymptotic
value of the tensor field-strength on the M5-brane
in the construction of the theory from M-theory.
The field $H$ is fluctuating and  is assumed to go to zero at infinity.

At order $O(\Th)^2$, the scalar fields have a quartic interaction:
\be\label{quartic}
-\frac{\pi}{2}\eta_{kn}\Th^{ijk}\Th^{lmn}
    \px{i}\Phi^I\px{j}\Phi^J\px{l}\Phi^I\px{m}\Phi^J.
\ee
These terms can be determined by dimensional reduction, as we 
will now explain.

\subsection{Dimensional reduction to 4+1D}
If we compactify on $\MS{1}$ of circumference $2\pi R$,
the dimensional reduction to 4+1D proceeds according to:
$$
F_{\u\v} = 2\pi R H_{\u\v 5},\qquad
\th^{\u\v} =\Th^{\u\v 5},\qquad
g^2 = 4\pi^2 R,\qquad
\phi^I = (2\pi)^{3/2} R\Phi^I.
$$
To first order in $\th$, the 4+1D action is \cite{KZ,GGS}:
\bear
L_{4+1D} &=&
\frac{1}{2g^2}\px{\u}\phi^I\qx{\u}\phi^I
+\frac{1}{4g^2}F_{\u\v}F^{\u\v}
\nn\\ &&
+\frac{1}{2g^2}\th^{\u\v}F_{\v\sigma}F^{\sigma\tau}F_{\tau\u}
-\frac{1}{2g^2}\th^{\u\v}F_{\u\v}F^{\sigma\tau}F_{\sigma\tau}
+\frac{1}{g^2}\th^{\u\v}F_{\u\sigma}\px{\v}\phi^I\qx{\sigma}\phi^I
-\frac{1}{4g^2}\th^{\u\v}F_{\u\v}\px{\sigma}\phi^I\qx{\sigma}\phi^I.
\nn
\eear
Assuming that supersymmetry protects the leading interactions
from loop-corrections, we can obtain (\ref{cubic})-(\ref{quartic})
by requiring that dimensional reduction should produce the latter
corrections in 4+1D SYM.

\section{Quantum-Mechanics on a blown-up $A_1$ singularity}\label{qmaone}
In this section we will describe in detail the quantum mechanics
on the target space -- the blown up $A_1$-singularity.
The quantum mechanics has $\SUSY{8}$ supersymmetry and this
is related to the hyper-K\"ahler structure on the target space.
For a generic description of QM on hyper-K\"ahler manifolds,
see \cite{HKLR}.

\subsection{The geometry}

The M(atrix)-model of $k$ coincident $M5$-branes in the
sector with longitudinal momentum $p_\parallel=N$ is
postulated \cite{ABS,NekSch,Berk}
to be described by Quantum Mechanics on the manifold
${\cal M} _{N,k}$ defined by
\begin{equation}
\pmatrix{
[X, X^{\dagger}] + [Y,Y^{\dagger}]
+ Z^{\dagger} Z - W^{\dagger} W &=& \xi^2 Id_{N \times N}
\cr
[X, Y] + Z^{\dagger} W &=& 0
}
/ U(N)
\label{MNk}
\end{equation}
where the group $U(N)$ acts on the $N\times N$ complex
matrices $X$ and $Y$ and on the $k\times N$ complex
matrices $Z$ and $W$ in the natural way.  For any
$N$ and $k$, the trace of matrices $X$ and $Y$ is
a flat four-dimentional space $\BR^4$ and so we can write
${\cal M}_{N,k} = \BR^4 \times \tilde {\cal M}_{N,k}$.
This flat four-dimensional part corresponds
to the center-of-mass coordinates.
We study a single M5-brane, $k=1$, and $N=2$.  The 
manifold $\tilde {\cal M}_{N=2,k=1} \equiv {\cal M}$ 
is the blown-up $A_1$ singularity.  It must be the same
as $\tilde{\cal M}_{N=1,k=2}$, as can be confirmed
explicitly.  The metric is easier to obtain in the second case, for
$k=2,N=1$, when $\cal M$ can be embedded in ${\BC}^4$ as
\begin{eqnarray}
\mbox{Tr} \left( A^{\dagger} A \sigma_i \right) = 
\xi^2 \delta^3_i  \qquad \qquad
A \sim e^{i \epsilon} A
\label{M}
\end{eqnarray}
where $A$ is a $2 \times 2$ complex matrix and
$\sigma_i$ are the Pauli matrices.

These conditions (ignoring the U(1) quotient for now) can be satisfied by parametrizing A as follows
\begin{eqnarray}
A = \xi  e^{i \epsilon} g 
\pmatrix {\cosh r& 0 \cr 0 &\sinh r}
\label{A}
\end{eqnarray}
where $g$ is an arbitrary element of SU(2), and $\xi$, $r$ and
$\epsilon$ are real.  The induced metric on this
five-(real)dimensional manifold is given by
\begin{eqnarray}
\frac { \mbox{Tr} \left( dA^{\dag} dA \right)}  {\xi^2} &=& 
\cosh(2 r) \left[ dr^2 + \frac {1} {2} \mbox{Tr}(dg^{\dag} dg)
\right] +
\frac {1} {\cosh (2 r)} \left [
(1\: 0) dg^{\dag} g
{1 \atopwithdelims ( ) 0} \right ]^2 
\nonumber \\ &+&
\cosh(2 r) \left[ d \epsilon 
+ \frac {i} {\cosh (2 r)} (1\: 0) dg^{\dag}g 
{1 \atopwithdelims ( ) 0}  \right]^2
\end{eqnarray}
To obtain the U(1) quotient, we must choose a function
$\epsilon(r,g)$ such that the distance between
$(r,g,\epsilon(r,g))$ and $(r+dr, g+dg, \epsilon(r+dr, g+dg))$ is
minimized.  This corresponds to choosing $\epsilon(r,g)$ in such a
way that the last term in the above equation vanishes.  Thus, the
metric on $\cal M$ is simply
\begin{equation}
\frac {ds^2 \vert _{\cal M}} {\xi^2} = 
\cosh(2 r) \left[dr^2 + \frac {1} {2} \mbox{Tr}
(dg^{\dag} dg) \right] +
\frac {1} {\cosh (2 r)}
\left[ (1 0) dg^{\dag} g
{1 \atopwithdelims ( ) 0} \right]^2
\end{equation}

It is apparent from equation (\ref{M}) that $\cal M$ is invariant
under SU(2) acting on A on the left and under U(1) acting on the
right.  $g$ can be parametrized by three angles $\theta$, $\phi$
and $\alpha$ in such a way as to make these symmetries explicit
\begin{equation}
g = 
\pmatrix{ \cos(\theta /2) e^{i(\alpha-\phi)/2} &
-\sin(\theta /2) e^{i(-\alpha-\phi)/2} \cr
\sin(\theta /2) e^{i(\alpha+\phi)/2} &
\cos(\theta /2) e^{i(-\alpha+\phi)/2} \cr}
\label{g}
\end{equation}
$\theta$ runs from 0 to $\pi$, and $\phi$ and $\alpha$ run from 0
to $2\pi$.  Under this parametrization we find that the metric
becomes
\begin{equation}
ds^2 = \xi^2 \left\{
\cosh (2r) dr^2 + \frac {\cosh(2r)} {4} (d\theta ^2 + 
\sin^2 \theta d\phi^2)
+ \frac {\cosh^2(2r)-1} {4 \cosh(2r)} (d\alpha - \cos \theta
d\phi)^2 \right\}
\end{equation}
It will be convenient to make the change of variables
$\cosh(2r)=R^2$ ($R$ runs from 1 to $\infty$)
\begin{equation}
ds^2 = \xi^2 \left\{
\frac {R^4} {R^4-1} dR^2 + 
\frac {R^2} {4} (d\theta ^2 + 
\sin^2 \theta d\phi^2)
+ \frac {R^4-1} {4 R^2} (d\alpha - \cos \theta
d\phi)^2
\right\}
\label{metric}
\end{equation}
It is apparent that, for large $R$, $\cal M$ approaches flat space,
namely $\BR^4/\BZ_2$.  This is the metric for the
blown-up $A_1$ singularity.

For later reference, let us here record that the scalar Laplacian is
\begin{eqnarray}
\bigtriangledown ^2 &=& 
\xi^{-2} \Biggl \{
\frac {1} {R^3} \partial _R 
\left( \frac {R^4-1}{R} \partial_R \right)
\nonumber \\ &+&
\frac {4} {R^2} \left[ \frac {1}{\sin \theta} \partial _{\theta}
\left(  \sin \theta \partial_{\theta} \right)
+ \frac {1} {\sin^2 \theta} \left(\partial ^2 _ {\phi} +
\partial ^2 _ {\alpha} + 2 \cos \theta
\partial _ {\phi} \partial _ {\alpha}\right) \right]
+ \frac {4} {R^2(R^4-1)} \partial ^2 _ {\alpha}
\Biggl \}
\label{lapl1}
\end{eqnarray}
and that it is fully separable.

It is known that the manifold $\cal M$ is a 
hyper-K\"ahler manifold, a fact crucial to the existence 
of ${\cal N} = 8$ supersymmetric QM.  An explicit 
hyper-K\"ahler structure for $\cal M$ will now be given.

The following coordinates
\begin{equation}
a \equiv (R^4-1)^{1/4}\sin \frac{\theta}{2}
\exp \left(\frac{i}{2} (\alpha+\phi)\right)
\qquad\qquad
b \equiv (R^4-1)^{1/4}\cos\frac{\theta}{2}
\exp \left(\frac{i}{2} (\alpha-\phi)\right)
\label{ab}
\end{equation}
are good complex coordinates for $\cal M$.  In these
coordinates, the metric is
\begin{eqnarray}
g_{a\bar a} &= g_{\bar a a} &= 
\frac {(a\bar a+b\bar b)^3 + b\bar b}
{(a\bar a+b\bar b)^2 \sqrt{(a\bar a+b\bar b)^2 + 1}}
\nonumber\\
g_{b\bar b} &= g_{\bar b b} &= 
\frac {(a\bar a+b\bar b)^3 + a\bar a}
{(a\bar a+b\bar b)^2 \sqrt{(a\bar a+b\bar b)^2 + 1}}
\nonumber\\
g_{a\bar b} &= g_{\bar b a} &= 
- \frac {\bar a b}
{(a\bar a+b\bar b)^2 \sqrt{(a\bar a+b\bar b)^2 + 1}}
\nonumber\\
g_{b\bar a} &= g_{\bar a b} &= 
- \frac {a\bar b}
{(a\bar a+b\bar b)^2 \sqrt{(a\bar a+b\bar b)^2 + 1}}
\end{eqnarray}
which is clearly Hermitian and easily confirmed to
be K\"ahler, with a K\"ahler potential
\begin{equation}
K = X - tan^{-1} X \qquad \mbox{where} \qquad 
X = \sqrt {(a\bar a + b\bar b)^2 + 1}
\label{K}
\end{equation}
Also, the determinant is $g = 1$, so $\cal M$ is Ricci-flat,
since $R_{i\bar j} = - \partial_i \partial_j \ln \det (g)$.
The K\"ahler form, $\Omega_1$, is as usual
$(\Omega_1)_{i \bar j} = -(\Omega_1)_{\bar j i}
= i g_{i \bar j}$.
We have 2 more K\"ahler forms $\Omega_2$ and $\Omega_3$,
satisfying the hyper-K\"ahler condition
$g^{\alpha\gamma} (\Omega_a)_{\alpha\beta}
(\Omega_b)_{\gamma\delta} = 
\epsilon_{a b c} (\Omega_c)_{\beta\delta}
+ \delta_{a b} g_{\beta\delta}$, where
$\alpha = i,\bar j$.
Explicitly, these are (written in
the complex coordinates $(a,b,\bar a, \bar b)$)
\begin{equation}
(\Omega_2)_{\alpha\beta} = 
\pmatrix{
0 & i & 0 & 0 \cr
-i & 0 & 0 & 0 \cr
0 & 0 & 0 & -i \cr
0 & 0 & i & 0\cr}  
\qquad
(\Omega_3)_{\alpha\beta} = 
\pmatrix{
0 & 1 & 0 & 0 \cr
-1 & 0 & 0 & 0 \cr
0 & 0 & 0 & 1 \cr
0 & 0 & -1 & 0\cr
}
\label{omega}
\end{equation}
Later, we will need a complex structure defined as usual by
\begin{equation}
(\Omega_2)_{\alpha\beta} = -g_{\alpha\gamma} J^{\gamma}_{\beta}
\label{J}
\end{equation}

\subsection{Quantization}

The construction of the ${\cal N} =8$ supersymmetric algebra and
its representation in a coordinate basis presented in this section
is general to any 2-(complex)dimensional hyper-K\"ahler manifold. 
We will first construct the ${\cal N} =4$ supersymmetric quantum
algebra, following a procedure similar to that in \cite{N=4}, and
then extent the supersymmetry to ${\cal N} =8$.

The Lagrangian is obtained by dimensionally reducing the d=4,
${\cal N} =1$ chiral supergravity Lagrangian in flat space
\cite{wb}.  Since after dimensional reduction the original SO(3,1)
spinor indices loose their meaning, we will use SU(2) spinors,
making no distinction between dotted and undotted spinor indices. 
Writing SU(2) spinors on the right and SO(3,1) spinors on the left
(using the spinor conventions in \cite{wb}) we define
\begin{equation}
\chi^{\alpha} \equiv \chi^{\alpha}
\qquad\qquad
\bar \chi^{\alpha} \equiv - \bar {\chi}_{\dot \alpha} = 
\bar {\sigma} ^{0 \dot \alpha \alpha} 
\bar {\chi}_{\dot \alpha}
\label{spinors1}
\end{equation}
The spinor products can be defined in our conventions as
\begin{equation}
\psi\chi\equiv \epsilon_{\alpha\beta}
\psi^{\alpha}\chi^{\beta} = \psi^{\alpha}\chi_{\alpha}
\qquad
\bar\psi\chi\equiv \epsilon_{\alpha\beta}
\bar\psi^{\alpha}\chi^{\beta} =
\bar\psi^{\alpha}\chi_{\alpha}
\qquad
\bar\psi \bar\chi \equiv \epsilon_{\alpha\beta}
\bar\psi^{\alpha} \bar\chi^{\beta}=
\bar\psi^{\alpha}\bar\chi_{\alpha}
\label{spinors2}
\end{equation}
Notice that this implies that
$(\chi^\alpha)^{\dagger}=\bar\chi_\alpha$ and
$(\chi_\alpha)^{\dagger}=-\bar\chi^\alpha$
which in turn gives
\begin{equation}
(\chi\phi)^{\dagger} = -\bar \chi \bar \phi
\qquad\qquad
(\bar \chi\bar\phi)^{\dagger} = -\chi\phi
\qquad\qquad
(\bar \chi\phi)^{\dagger} = \bar \chi  \phi
\label{conjugate}
\end{equation}
Finally, following \cite{wb}, 
$\epsilon^{12} = -\epsilon_{12} = 1$.

In this notation, the dimensionally reduced Lagrangian is
\begin{eqnarray}
{\cal L} &=& g_{i \bar j} \dot z^i \dot {\bar z} ^{\bar j}
- i g_{i \bar j} \bar \chi ^{\bar j} (D_t \chi^i)
- \frac {1} {4} R_{i \bar j k \bar l}(\chi^i \chi^k)
(\bar \chi ^{\bar j} \bar \chi ^{\bar l})
\nonumber \\ &\sim&
g_{i \bar j} \dot z^i \dot {\bar z} ^{\bar j}
- \frac {i}{2} g_{i \bar j} \bar \chi ^{\bar j}
(D_t \chi^i)
+ \frac {i}{2} g_{i \bar j} (D_t \bar \chi ^{\bar j})
\chi^i
- \frac {1} {4} R_{i \bar j k \bar l}(\chi^i \chi^k)
(\bar \chi ^{\bar j} \bar \chi ^{\bar l})
\label{L}
\end{eqnarray}
where
$D_t \chi^i = \dot \chi + \Gamma ^i_{ab}\chi^a\dot z^b$
and the two forms of the Lagrangian differ only by a total time
derivative.  We will use the second form, which is real.

The conjugate momenta are
\begin{eqnarray}
P_i &\equiv&  \frac {\partial {\cal L}}{\partial z^i} = 
g_{i\bar j} \dot {\bar z}^{\bar j} - \frac{i}{2}
\partial_k g_{i\bar j} \bar \chi^{\bar j} \chi^k
\nonumber \\
\bar P_{\bar j} &\equiv& \frac {\partial {\cal L}}{\partial \bar
z^{\bar j}} = 
g_{i\bar j} \dot z^i + \frac{i}{2}
\partial_{\bar l} g_{i\bar j} \bar \chi^{\bar l} \chi^i
\nonumber \\
\pi_{i\alpha} &\equiv&  \frac {\partial {\cal L}}{\partial
\chi^{i\alpha}} = 
- \frac {i} {2} g_{i\bar j} \bar \chi ^{\bar j} _{\alpha}
\nonumber \\
\bar \pi_{\bar j \alpha} &\equiv&  \frac {\partial {\cal
L}}{\partial
\bar \chi^{\bar j\alpha}} = 
+ \frac {i} {2} g_{i\bar j} \chi ^i _{\alpha}
\label{conjmom}
\end{eqnarray}
And their canonical Poisson brackets are
\begin{eqnarray}
\{ z^k, P_i\} &= \delta^i_k \qquad \qquad\qquad \quad
\{ \bar z^{\bar j}, \bar P_{\bar l} \} &= \delta^{\bar j}_{\bar l}
\nonumber \\
\{\chi^{i\alpha}, \pi_{k \beta} \} &= -\delta^i_k
\delta^{\alpha}_{\beta}
\qquad \qquad
\{\bar\chi^{\bar j \alpha}, \bar \pi_{\bar l \beta} \} 
&= -\delta^{\bar j}_{\bar l} \delta^{\alpha}_{\beta}
\label{poisson}
\end{eqnarray}
This system possesses fermionic constraints of the second kind
\begin{equation}
\phi_i^\alpha \equiv \pi_i^\alpha + \frac{i}{2}g_{i\bar j}
\bar \chi ^{\bar j \alpha}=0 \qquad\qquad
\bar \phi_{\bar j \alpha} \equiv \bar \pi_{\bar j\alpha}
- \frac{i}{2}g_{i\bar j} \chi ^i_\alpha=0
\end{equation}
whose Poisson bracket is
\begin{equation}
\{ \phi_i^\alpha, \bar\phi_{\bar j \beta}\} = 
-ig_{i\bar j} \delta ^\alpha_\beta
\end{equation}
Following standard procedure \cite{casalbuoni}, we define the Dirac
brackets
\begin{equation}
\{ \circ, \circ\}_D \equiv 
\{ \circ, \circ\}
-
\{ \circ, \phi_i^\alpha \}
\frac {1} { \{ \phi_i^\alpha, \bar\phi_{\bar j \beta}\}}
\{ \bar\phi_{\bar j \beta}, \circ \}
-
\{ \circ, \bar\phi_{\bar j \beta}\}
\frac {1} { \{\bar\phi_{\bar j \beta}, \phi_i^\alpha \}}
\{\phi_i^\alpha , \circ \}
\end{equation}
Canonical quantization proceeds by substituting
$\{\circ,\circ\}_D$ $\to$ $-i\lbrack\circ, \circ\rbrack$
or  $-i\{\circ, \circ\}$, as appropriate.  Evaluating the Dirac
brackets for all quantities, we obtain the following algebra
\begin{equation}
\lbrack z^k, P_i \rbrack = i \delta^k_i
\qquad
\lbrack \bar z^{\bar l}, \bar P_{\bar j} \rbrack = i
\delta^{\bar l}_{\bar j}
\qquad
\{ \chi^{i\alpha}, \bar \chi ^{\bar j}_{\beta} \} = 
g^{i \bar j} \delta^{\alpha}_{\beta}
\label{comm1}
\end{equation}
The commutators of $z_i$ and $\bar z_{\bar j}$ with
the fermions $\chi^{i\alpha}$ and 
$\bar \chi ^{\bar j \beta}$ vanish.
Define
\begin{equation}
K_i \equiv \frac{i}{2} \partial_k g_{i\bar j}
\bar \chi^{\bar j \beta} \chi^k_{\beta}
\qquad \qquad
\bar K_{\bar j} \equiv -\frac{i}{2} \partial_{\bar l}
g_{i\bar j} \bar \chi^{\bar l \beta} \chi^i_{\beta}
\end{equation}
The commutation relationships of $P_k$ and $\bar P_{j}$ 
alone are not as relevant to what will follow as the commutation
relationships involving $P+K$ and 
$\bar P+ \bar K$.
With some work, it can be shown that
\begin{eqnarray}
\lbrack (P+K)_k, \chi^{i \alpha} \rbrack &= 
i \Gamma^i_{ka}\chi^{a \alpha}
\qquad \qquad
\lbrack (P+K)_k, \bar \chi^{\bar j \alpha} \rbrack &= 0
\nonumber  \\
\lbrack (\bar P+\bar K)_{\bar l}, 
\chi^{i \alpha} \rbrack &= 0
\qquad \qquad \qquad \qquad
\lbrack (\bar P+\bar K)_{\bar l}, 
\bar \chi^{\bar j \alpha} \rbrack &=
i {\bar\Gamma}^{\bar j}_{\bar l \bar b}
\bar \chi^{\bar b \alpha}
\label{comm2}
\end{eqnarray}
and that
\begin{equation}
\lbrack (P+K)_i, (P+K)_k \rbrack = 
\lbrack (\bar P+\bar K)_{\bar j},
(\bar P+\bar K)_{\bar l} \rbrack = 0
\qquad 
\lbrack (P+K)_i, (\bar P + \bar K)_{\bar j} \rbrack =
- R_{i \bar j a \bar b} \chi^a \bar \chi^{\bar b}
\label{comm3}
\end{equation}
The (classical) Hamiltonian obtained by a standard procedure is
\begin{equation}
H = g^{i \bar j} (P+K)_i (\bar P+\bar K)_{\bar j} +
\frac {1}{4} R_{i \bar j a \bar b} 
(\chi^i \chi^a) (\bar \chi^{\bar j} \bar \chi^{\bar b})
\label{H}
\end{equation}
Since $R _{a \bar b} = g^{i\bar j}  R_{i \bar j a \bar b} = 0$,
there are no ordering ambiguities and thus
(\ref{H}) can be simply taken to be the quantum Hamiltonian.
A very important cyclic identity for the fermionic coordinates is
that
\begin{equation}
\chi^{\alpha} (\psi \phi) + \psi^{\alpha}(\phi\chi) +
\phi^{\alpha}(\chi\psi) = 0
\end{equation}
This identity must be used with care for the quantum fermionic
coordinates, since they have non-trivial anti-commutation
relationships.  Nevertheless, the Hamiltonian can be rewritten as
\begin{equation}
H = g^{i \bar j} (p+K)_i (\bar p+\bar K)_{\bar j} -
\frac {1}{2} R_{i \bar j a \bar b} 
(\chi^i \bar \chi^{\bar j}) (\chi^a \bar \chi^{\bar b})
\end{equation}
The (classical) supersymmetry transformations are given in
\cite{wb}
\begin{equation}
\delta_{\xi} z^i = \xi \chi^i
\qquad \qquad
\delta_{\xi} \chi^{i\alpha} = 
-i \bar \xi^{\alpha} \dot z^i
-\Gamma^i_{jk}(\xi\chi^j)\chi^{k\alpha}
\end{equation}
With some work, the N\"oether current can be calculated for the
real form of the Lagrangian (\ref{L}) to be
\begin{equation}
J_{SUSY} = g_{i\bar j} \dot{\bar z}^{\bar j} \xi \chi^i - 
g_{i\bar j} \dot z^i \bar \xi \bar \chi^{\bar j}
\end{equation}
The minus sign is related to the minus sign in equation
(\ref{conjugate}).  Defining the SUSY generators through
$J_{SUSY} = Q^{\alpha} \xi_{\alpha} 
+ \bar Q_{\alpha} \bar \xi^{\alpha}$
we obtain that
\begin{equation}
Q^{\alpha} = 
g_{i\bar j} \dot{\bar z}^{\bar j} \chi^{i\alpha} =
(P+K)_i\chi^{i\alpha}
\qquad\qquad
\bar Q_{\alpha} = 
g_{i\bar j} \dot z^i \bar \xi \bar \chi^{\bar j} =
(\bar P + \bar K)_{\bar j} \bar \chi ^{\bar j}_{\alpha}
\end{equation}
There are no ordering ambiguities when interpreting these as
quantum operators.

It is now a matter of another careful computation to confirm that
the Q's satisfy the SUSY algebra
\begin{equation}
\{Q^{\alpha}, Q^{\beta}\} = 
\{\bar Q_{\alpha}, \bar Q_{\beta}\} = 0
\qquad \qquad  \{Q^{\alpha}, \bar Q_{\beta}\} = 
\delta^\alpha_{\beta} H
\end{equation}

We must represent this algebra in a coordinate basis.  There are 4
fermionic coordinates and thus the Hilbert space can be written as
a $2^4 = 16$-dimensional vector of complex functions of $z$ and
$\bar z$.  Equation (\ref{comm1}) suggests that we associate $P_i$
with $-i\partial_i$ and 
$\bar P_{\bar j}$ with $-i\partial_{\bar j}$.  Naively, this would
disagree with (\ref{comm2}) and (\ref{comm3}).  We need to pay
attention to how these partial derivatives act on the fermionic
basis.

Let us begin by noticing that
\begin{equation}
\lbrack (P+K)_k, \chi_{\bar j}^{\alpha} \rbrack =
\lbrack (P+K)_k, \bar \chi^{\bar j \alpha} \rbrack = 0
\qquad\qquad
\{ \chi_{\bar l}^{\alpha}, \bar\chi^{\bar j}_{\beta} \} =
\delta^{\bar j}_{\bar l} \delta^{\alpha}_{\beta}
\label{pk}
\end{equation}
and
\begin{equation}
\lbrack (\bar P+\bar K)_{\bar l}, \chi^{i\alpha} \rbrack =
\lbrack (\bar P+\bar K)_{\bar l}, \bar \chi_{i}^{\alpha}
 \rbrack = 0
\qquad\qquad
\{ \chi^{i\alpha}, \bar\chi_{k\beta} \}
= \delta^i_k \delta^{\alpha}_{\beta}
\label{pkbar}
\end{equation}
\def \vacd {|\!\downarrow \rangle}
\def \vacu {|\!\uparrow \rangle}
Define $\vacd$ by 
$\bar \chi^{\bar j}_{\beta} \vacd =
\bar \chi_{i\beta} \vacd = 0$.
It is consistent with all commutation relationships to write that 
$(P+K)_i \lbrack f(z,\bar z) \vacd \rbrack = 
(-i \partial_i f(z,\bar z)) \vacd$ and
$(\bar P+\bar K)_{\bar j} \lbrack 
f(z,\bar z) \vacd \rbrack = 
-i \partial_{\bar j} f(z,\bar z) \vacd$
>From the no-fermions state $\vacd$ we can construct the one-fermion
state in two ways: using $\chi^{i\alpha}$ or $\chi_{\bar
j}^{\alpha}$.  Let 
$|\chi^{i\alpha}\rangle \equiv \chi^{i\alpha}\vacd$ and
$|\chi_{\bar j}^{\alpha}\rangle \equiv 
\chi_{\bar j}^{\alpha} \vacd =  
g_{i\bar j}\chi^{i\alpha}\vacd$. 
It can be then checked by an explicit calculation that the
following is consistent with all commutation relationships
\begin{eqnarray}
(P+K)_i \lbrack f^{\bar j}_{\alpha}(z,\bar z) 
|\chi_{\bar j}^{\alpha}\rangle \rbrack = 
(-i \partial_i f^{\bar j}_{\alpha}(z,\bar z))
|\chi_{\bar j}^{\alpha}\rangle
\nonumber \\
(\bar P+\bar K)_{\bar j} \lbrack f_{i\alpha}(z,\bar z) 
|\chi^{i\alpha}\rangle \rbrack =
(-i \partial_{\bar j} f_{i\alpha}(z,\bar z))
|\chi^{i\alpha}\rangle
\end{eqnarray}
Thus, one should think of $\vacd$ as being independent of both $z$
and $\bar z$, of $\chi^{i\alpha}$ and 
$\bar\chi_i^{\alpha}$ as being holomorphic functions and 
$\chi_{\bar j}^{\alpha}$ and $\bar\chi^{\bar j\alpha}$ as being
antiholomorphic functions.

We can continue this procedure by defining two, three and
four-fermion states $|\chi\chi\rangle$, $|\chi\chi\chi\rangle$ and
$|\chi\chi\chi\chi\rangle$, and checking each time that the
commutation relationships work. The computation is made easier if
one notices that the four-fermion state $|\chi\chi\chi\chi\rangle$
is the same as $\vacu$ defined by
$\chi_{\bar j}^{\beta} \vacu = \chi^{i\beta} \vacu = 0$
and that $|\chi\chi\chi\rangle$ can be written as $|\bar\chi\rangle
\equiv \bar \chi \vacu$.
With this explicit construction, we can write the energy eigenvalue
equation $H|\rangle = E|\rangle$ as a differential equation. 
Notice that H commutes with the fermion number operator $\chi^i\bar
\chi_i$ and so we can write different differential equations for
each fermion number.
For the no-fermions state, $f\vacd$, 
$H|\rangle = E|\rangle$ can be written as
\begin{equation}
-g^{i\bar j} \partial_i \partial_{\bar j} f(z,\bar z)
= E f(z,\bar z)
\label{Hf=Ef0}
\end{equation}
For the one-fermion states, 
$f_{i\alpha}(z,\bar z)|\chi^{i\alpha}\rangle$,
the equation is
\begin{equation}
-g^{k\bar j} \partial_k \partial_{\bar j} f_{i\alpha} 
+g^{k\bar j}\Gamma^a_{ik}\partial_{\bar j} f_{a\alpha}
= E f_{i\alpha}
\label{Hf=Ef1}
\end{equation}
And, finally, for the two-fermion states, 
$f_{nm\alpha\beta}(z,\bar z)
|\chi^{n\alpha}\chi^{m\beta}\rangle$,
we obtain
\begin{equation}
-g^{k\bar j} \partial_k \partial_{\bar j} 
f_{nm\alpha\beta} 
+g^{k\bar j}
(\Gamma^a_{nk}\partial_{\bar j} f_{am\alpha\beta} +
\Gamma^a_{mk}\partial_{\bar j} f_{na\alpha\beta})
-R_{n\bar b m\bar d} g^{a\bar b} g^{c\bar d} 
f_{ac\alpha\beta}
= E f_{nm\alpha\beta}
\label{Hf=Ef2}
\end{equation}
The equations for three- and four-fermion states can be obtained by
analogy to (\ref{Hf=Ef1}) and (\ref{Hf=Ef0}).

Given the tensor $J$ defined in equation (\ref{J}), we obtain four
more supersymmetry generators
\begin{equation}
S^{\alpha} \equiv
(\bar P + \bar K)_{\bar j} J^{\bar j}_{\;i} \chi^{i\alpha}
\qquad\qquad
\bar S_{\alpha} \equiv
(P+K)_i J^i_{\; \bar j} \bar \chi ^{\bar j}_{\alpha}
\end{equation}
These can be confirmed to satisfy ${\cal N} = 8$ SUSY algebra,
namely
\begin{equation}
\{Q^\alpha, \bar Q_\beta\} = \{S^\alpha, S_\beta\} = 
\delta^\alpha_\beta H
\end{equation}
and
\begin{eqnarray}
\{Q^\alpha, \bar S_\beta\} =
\{S^\alpha, \bar Q_\beta\} =
\{Q^\alpha, Q^\beta\} =
\{S^\alpha, Q^\beta\} &=&
\nonumber\\
\{S^\alpha, S^\beta\} =
\{\bar Q_\alpha, \bar Q_\beta\} =
\{\bar S_\alpha, \bar Q_\beta\} =
\{\bar S_\alpha, \bar S_\beta\} &=& 0
\label{N=8}
\end{eqnarray}
In proving the above relationships, we need to use the properties
of $J$: the metric is hermitian, $J$ is covariantly constant and
the Nijenhuis tensor, ${\cal N} (J)$, vanishes.

\subsection{Scattering}\label{scat}

Consider the simplest scattering question -
 s-wave scattering of one of the
scalar particles in the low energy regime.  This is described by
$-\nabla^2 f = E f$.  The Laplacian was given in equation
(\ref{lapl1}).
 For s-wave scattering,
$f = f(R)$ and the differential equation to be solved is
\begin{equation}
-\frac {1} {R^3} \partial _R \left(\frac {R^4-1}{R} \partial_R f(R)
\right)
= \xi^2 E f(R)
\label{eq} 
\end{equation}
Let $y^4 = \xi^2 E \ll 1$ and define $x = y R$.  We can then
rewrite (\ref{eq}) as
\begin{equation}
-\frac {1} {x^3} \partial _x \left(\frac {x^4-y^4}{x} \partial_x
f(x) \right)
= y^2 f(x)
\label{eq1}
\end{equation}

For $x \ll 1/y$ we can treat the RHS of equation (\ref{eq1}) as a
small perturbation.  Keeping the solution finite at x=y, we obtain
a solution as an expansion in y
\begin{equation}
1 - \frac {1}{8} x^2 y^2 + \frac {1}{192} x^4 y^4
- \frac {1}{9216} x^6 y^6 + o(y^8)
\label{A1}
\end{equation}

For $x \gg y$, rewrite (\ref{eq1}) as
\begin{equation}
-\frac {1} {x^3} \partial _x \left(x^3
\partial_x f(x) \right)
- y^2 f(x) = 
- y^4 \frac {1} {x^3} \partial _x \left(\frac {1}{x}
\partial_x f(x) \right)
\label{eq2}
\end{equation}
The RHS is again a small perturbation.  The solution with the RHS
set to zero is the asymptotic solution at infinity
\begin {equation}
\frac {2}{y} \frac {J_1(y x) + a Y_1(y x)} {x}
\end{equation}
(the factor $(2/y)$ is for convenience only).  
We need to obtain the lowest order correction due to the RHS, which
is $-y^6/24 x^2$.  The expansion in y is thus
\begin{equation}
1 - \frac {1}{8} x^2 y^2 + \frac {1}{192} x^4 y^4
- \left(\frac {1}{9216} x^6 + \frac {1} {24 x^2} \right)
y^6 + o(y^8)
- \frac {4 a} {\pi y^2 x^2} + ...
\label{A2}
\end{equation}
Since the region of validity of the two expansions (\ref{A1}) and
(\ref{A2}) in y overlaps for $x \sim 1$, they should match. 
We need to cancel the $x^{-2}$ terms in expansion (\ref{A2}), and
thus
\begin{equation}
a = - \frac {\pi y^8} {96} = - \frac {\pi \xi^4 E^2} {96}
\label{a}
\end{equation}

\subsection{The $SO(5)$ symmetry and two-particle states}

The tensor multiplet on the M5-brane contains 5 real
scalars with an $SO(5)$ symmetry.  This symmetry should
somehow be visible in the M(atrix)-model we have just 
constructed.  In this section we will find the $SO(5)$
symmetry which commutes with the Hamiltonian and
construct a real five-dimentional representation of it.

The $SO(5)$ symmetry we are looking for contains the
$SO(3)_{\Omega}$ symmetry acting naturally on the three 
K\"ahler forms 
$\Omega_i$'s, as well as the $SU(2)=SO(3)_{f}$ symmetry acting
on the fermionic indices.  Since the $SO(5)$ symmetry is 
connected to the scalars, it should act non-trivially on
the states $f\vacd$, $f\vacu$ and  
$|\chi^{n\alpha}\chi^{m\beta}\rangle$.
Let us see what is the action of the $SO(3)_{\Omega}$
symmetry on the fermion vacuum states.  
For example, consider using
$\Omega_1 + \epsilon \Omega_2$
instead of $\Omega_1$ to define the complex coordinates
$(z^i,\bar z^{\bar j})$.
The required change of coordinates is
\begin{equation}
dz^i \rightarrow dz^i + \epsilon g^{i\bar j}
(\Omega_2)_{\bar j \bar k} d \bar z^{\bar k}
\qquad\qquad
d \bar z^{\bar i} \rightarrow d\bar z^{\bar i} + 
\epsilon g^{\bar  i j} (\Omega_2)_{j k} dz^k
\end{equation}
This being just a change of variables, we know how
it will act on the fermionic variables $\chi^{i\alpha}$
\begin{equation}
\chi^{i\alpha} \rightarrow \chi^{i\alpha} 
+ \epsilon g^{i\bar j}
(\Omega_2)_{\bar j \bar k} \bar\chi^{\bar k}_{\alpha}
\qquad\qquad
\bar \chi^{\bar i}_\alpha \rightarrow 
\bar \chi^{\bar i}_\alpha + 
\epsilon g^{\bar  i j} (\Omega_2)_{j k} \chi^{k\alpha}
\end{equation}
The new $\bar \chi$'s must annihilate the new vacuum,
and so we infer that
\begin{equation}
\vacd \rightarrow \vacd
- \frac{1}{2} \epsilon (\Omega_2)_{ij} 
\left (\chi^{i1}\chi^{j1}+\chi^{i2}\chi^{j2} \right ) \vacd
\end{equation}
where $1,2$ are concrete fermionic indices.
Similarly, we obtain the action on $\vacu$
\begin{equation}
\vacu \rightarrow \vacu
- \frac{1}{2} \epsilon (\Omega_2)_{\bar i \bar j} 
\left (\bar \chi^{\bar i 1}\bar \chi^{\bar j 1}
+\bar \chi^{\bar i 2}\bar \chi^{\bar j 2} \right ) \vacu
\end{equation}
which can be rewritten as
\begin{equation}
\vacu \rightarrow \vacu
- \frac{1}{2} \epsilon (\Omega_2)_{ij} 
\left (\chi^{i1}\chi^{j1}+\chi^{i2}\chi^{j2} \right ) \vacd
\end{equation}
if we define that
\begin{equation}
\vacu \equiv \chi^{11}\chi^{21}\chi^{12}\chi^{22} \vacd
\end{equation}
We also need
\begin{equation}
(\Omega_2)_{ij} 
\left (\chi^{i1}\chi^{j1}+\chi^{i2}\chi^{j2} \right ) \vacd
\rightarrow
(\Omega_2)_{ij} 
\left (\chi^{i1}\chi^{j1}+\chi^{i2}\chi^{j2} \right ) \vacd
+ 4 \epsilon \left ( \vacu + \vacd \right )
\end{equation}
Repeating this calculation for the change of coordinates
generated by using $\Omega_1 + \epsilon \Omega_3$
instead of $\Omega_1$,we obtain that
the following three
states form a basis for a real three dimentional
representation of $SO(3)_\Omega$
\begin{eqnarray}
\frac {1}{\sqrt{2}} \left ( \vacu + \vacd  \right )
\nonumber\\
\frac {1}{2 \sqrt {2}} 
(\Omega_2)_{ij} 
\left (\chi^{i1}\chi^{j1}+\chi^{i2}\chi^{j2} \right ) \vacd
\nonumber\\
\frac {i}{\sqrt{2}} \left ( \vacd - \vacu  \right )
\end{eqnarray}
The first and third states are singlets under $SO(3)_f$,
but the second one is not - it is part of a triplet.  
Filling in the triplet gives us the complete basis for 
the $SO(5)$ representation we were after
\begin{eqnarray}
|\phi^0 \rangle &\equiv&
\frac {1}{\sqrt{2}} \left ( \vacu + \vacd  \right )
\nonumber\\
|\phi^1 \rangle &\equiv&
\frac {i}{2 \sqrt {2}} 
(\Omega_2)_{ij} 
\left (\chi^{i1}\chi^{j1}-\chi^{i2}\chi^{j2} \right ) \vacd
\nonumber\\
|\phi^2 \rangle &\equiv&
\frac {1}{2 \sqrt {2}} 
(\Omega_2)_{ij} 
\left (\chi^{i1}\chi^{j1}+\chi^{i2}\chi^{j2} \right ) \vacd
\nonumber\\
|\phi^3 \rangle &\equiv&
\frac {-i}{2 \sqrt {2}} 
(\Omega_2)_{ij} 
\left (\chi^{i1}\chi^{j2}+\chi^{i2}\chi^{j1} \right ) \vacd
\nonumber\\
|\phi^4 \rangle &\equiv&
\frac {i}{\sqrt{2}} \left ( \vacd - \vacu  \right )
\end{eqnarray}
By construction the $SO(5)$ transformations must commute 
with the Hamiltonian.
This manifests itself in the fact that
\begin{equation}
H \left ( f_i(z,\bar z) |\phi^i \rangle \right ) = 
\left (-\nabla^2 f_i(z,\bar z) \right ) |\phi^i \rangle 
\end{equation} 
as can be checked explicitly.  All five of the
scalars have the same equation of motion.
We now know which of the 6 two-fermion states
$|\chi\chi\rangle$ correspond to scalars and which to
the 3-form field.  The later is represented by the singlet
states under $SO(3)_f$, namely by 
$f_{ij}\epsilon_{\alpha\beta}|\chi^{i\alpha} 
\chi^{j\beta}\rangle$
where $f_{ij}=f_{ji}$ has three independent components 
as needed.

\subsection{Relation to the M(atrix)-model}
The calculation that we performed should be compared with
the $O(\Th)^2$ scattering in the M(atrix)-model of the
noncommutative tensor multiplet. We will not perform the comparison
in detail here \cite{WIP}.
For the scattering of two massless particles, there are 4
Feynman diagrams that contribute at $O(\Th)^2$.
One from the quartic vertex (\ref{quartic}) and
the other contributions are from the tree diagrams
(the $s,t,u$ channels) with two cubic $\Phi\Phi H$ vertices
as in (\ref{cubic}) and an $\lbrack H H\rbrack$ propagator.

Let us see what does this all mean for scattering of 
two scalar particles on the M5-brane.  Let the fermionic 
coordinates corresponding to the centre-of-mass 
coordinates be denoted by $\psi^{i\alpha}$ , and those 
corresponding to the individual particles be 
$A^{i\alpha} = (\chi^{i\alpha} + \psi^{i\alpha})/2$
and
$B^{i\alpha} = (\chi^{i\alpha} - \psi^{i\alpha})/2$.
Since for scattering we are only interested in
asymptotic states, we can assume that everything
happens on a flat manifold.  Now, a state of two
identical scalars $A$ and $B$ is just 
$|\phi^0\rangle_A |\phi^0\rangle_B$
where we chose a specific direction under the
SO(5).  This can be rewritten in terms of
the separated coordinates $\chi$ and $\phi$
\begin{eqnarray}
|\phi^0\rangle_A |\phi^0\rangle_B &=& 
\frac {1}{2} \left [
\vacd_A\vacd_B + \vacu_A\vacu_B +
(\vacd_A\vacu_B + \vacd_A\vacu_B)
\right ] 
\nonumber\\
&=& 
\frac {1}{2} \left (
\vacd_\phi\vacd_\chi + \vacu_\phi\vacu_\chi
\right ) +
\sum |\rangle_\phi |\mbox {scalar} \rangle_\chi +
\sum |\rangle_\phi |\mbox {3-form} \rangle_\chi 
\end{eqnarray}
Choosing another-two particle state, we obtain
\begin{eqnarray}
|\phi^4\rangle_A |\phi^4\rangle_B &=& 
-\frac {1}{2} \left [
\vacd_A\vacd_B + \vacu_A\vacu_B -
(\vacd_A\vacu_B + \vacd_A\vacu_B)
\right ] 
\nonumber\\
&=& 
-\frac {1}{2} \left (
\vacd_\phi\vacd_\chi + \vacu_\phi\vacu_\chi
\right ) +
\sum |\rangle_\phi |\mbox {scalar} \rangle_\chi +
\sum |\rangle_\phi |\mbox {3-form} \rangle_\chi 
\end{eqnarray}
Thus, the scattering matrix for 
$\psi^I\psi^I \rightarrow \psi^J\psi^J$
will have a form
\begin{equation}
{}_A\langle\phi^4| \qquad {}_B\langle\phi^4| \quad S \quad
|\phi^0\rangle_A \qquad |\phi^0\rangle_B =
-\frac{1}{2}\sigma_1 + \sigma_2 +\sigma_3
\end{equation}
and that for $\psi^I\psi^I \rightarrow \psi^I\psi^I$
will be
\begin{equation}
{}_A\langle\phi^4| \qquad {}_B\langle\phi^4|\quad S\quad
|\phi^0\rangle_A\qquad |\phi^0\rangle_B =
\frac{1}{2}\sigma_1 + \sigma_2 +\sigma_3
\end{equation}
where
\begin{eqnarray}
\sigma_1 &=& 
{}_{\chi}\langle\downarrow| S \vacd_\chi =
{}_{\chi}\langle\uparrow| S \vacu_\chi
\nonumber \\
\sigma_2 &=& 
\left (
\sum {}_\chi\langle\mbox{scalar}|
\right ) S \left (
\sum |\mbox {scalar} \rangle_\chi 
\right )
\nonumber \\
\sigma_3 &=& 
\left (
\sum {}_\chi\langle\mbox{3-form}|
\right ) S \left (
\sum |\mbox {3-form} \rangle_\chi 
\right )
\end{eqnarray}
where we have ignored the (trivial) evolution
of the centre-of-mass coordinates.

The answers are as we would expect: the difference
(equal to $\sigma_1$) between the matrix element for 
$\psi^I\psi^I \rightarrow \psi^I\psi^I$
and for
$\psi^I\psi^I \rightarrow \psi^J\psi^J$
arises to the lowest order in the effective theory 
from the extra four-scalar vertex 
$\psi^I\psi^I\psi^J\psi^J$ which is zero
for $I=J$.  The other two pieces,
$\sigma_2$ and $\sigma_3$ will have different
momentum behaviour, as can be seen in an explicit
Feynmann diagram computation.

It is interesting to see what happens in the approximation 
of a large impact parameter. From the M(atrix)-model we
expect a force that behaves as $v^2/r^4$ where $r$ is the distance
between the particles and $v$ is the relative transverse velocity.
In the large impact parameter approximation, the $t$-channel
dominates. From (\ref{Gp}) we see that the $H$-propagator
behaves as $\frac{1}{r^4}$ when there is no longitudinal
momentum transfer. The two $\Phi\Phi H$ vertices should contribute
a $v^2$.


\section*{Acknowledgments}
We are very grateful to M. Berkooz for discussions.
We also wish to thank O. Aharony and A. Volovich for comments on
a previous version.
The research of OJG is supported by NSF grant number PHY-9802498.
The research of 
JLK is in part supported by NSERC (Natural Sciences and Engineering
Research Council of Canada).

\appendix

\section{Scattering in the field theory}
In this appendix we will describe how to calculate the scattering
in field theory.
Let us consider the scattering of two scalars to lowest nontrivial order.
There are two contributions.
One from the $O(\Th)^2$ quartic vertex and one from
a tree diagram with an $H$-field exchanged.
Let us consider the amplitude
$$
A^{I_1 I_2 I_3 I_4}(p_1, p_2, p_3, p_4),
$$
with $p_1+p_2 = p_3 + p_4$.

The amplitude is a sum of $s,t,u$ channels and a quartic vertex:
$$
A = A_t + A_u + A_s + A_q.
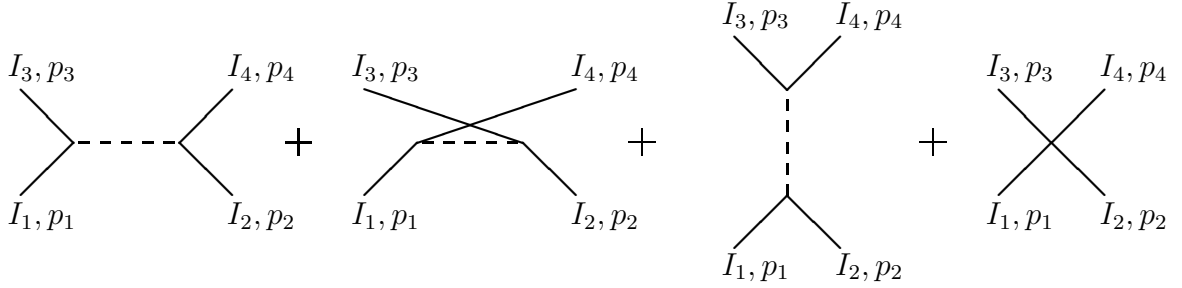
$$

\vskip 0.5cm
\parbox[c]{100mm}{
\begin{picture}(500,100)

\thicklines
\multiput(32,50)(8,0){5}{\line(1,0){4}}
\put(30,50){\line(-1,1){20}}
\put(30,50){\line(-1,-1){20}}
\put(70,50){\line(1,1){20}}
\put(70,50){\line(1,-1){20}}

\put(110,50){\line(1,0){10}}
\put(115,45){\line(0,1){10}}

\put(5,19){$I_1,p_1$}
\put(88,19){$I_2,p_2$}
\put(5,75){$I_3,p_3$}
\put(88,75){$I_4,p_4$}

\thicklines
\multiput(162,50)(8,0){5}{\line(1,0){4}}
\put(160,50){\line(3,1){60}}
\put(160,50){\line(-1,-1){20}}
\put(200,50){\line(-3,1){60}}
\put(200,50){\line(1,-1){20}}

\thinlines
\put(240,50){\line(1,0){10}}
\put(245,45){\line(0,1){10}}

\put(135,19){$I_1,p_1$}
\put(218,19){$I_2,p_2$}
\put(135,75){$I_3,p_3$}
\put(218,75){$I_4,p_4$}

\thicklines
\multiput(300,30)(0,8){5}{\line(0,1){4}}
\put(300,30){\line(-1,-1){20}}
\put(300,30){\line(1,-1){20}}
\put(300,70){\line(-1,1){20}}
\put(300,70){\line(1,1){20}}

\thinlines
\put(350,50){\line(1,0){10}}
\put(355,45){\line(0,1){10}}

\put(275,0){$I_1,p_1$}
\put(318,0){$I_2,p_2$}
\put(275,95){$I_3,p_3$}
\put(318,95){$I_4,p_4$}

\thicklines
\put(400,50){\line(1,1){20}}
\put(400,50){\line(1,-1){20}}
\put(400,50){\line(-1,1){20}}
\put(400,50){\line(-1,-1){20}}

\put(375,19){$I_1,p_1$}
\put(418,19){$I_2,p_2$}
\put(375,75){$I_3,p_3$}
\put(418,75){$I_4,p_4$}

\end{picture}}

\subsection{The Feynman rules}
The Feynman rules are as follows.

\vskip 0.5cm
\parbox[c]{100mm}{
\begin{picture}(500,150)

\thicklines
\multiput(20,50)(8,0){5}{\line(1,0){4}}

\thinlines
\put(90,52){\line(1,0){10}}
\put(90,48){\line(1,0){10}}

\put(0,45){$H_{ijk}$}
\put(60,45){$H_{lmn}$}
\put(38,60){$p$}

\put(120,47){$
{{36\pi}\over {p^2 + i\epsilon}}
  \eta_{rl}\eta_{sm}\eta_{tn}
  \left(p_{\lbrack i} p^{\lbrack r}
    \delta_j^{s}\delta_{k\rbrack}^{t\rbrack}
  -\frac{1}{6}  p_u{\epsilon_{ijk}}^{u\lbrack st} p^{r\rbrack}
\right)
$}
\put(130,25){$
+{{\pi}\over {2}}\left(
\epsilon_{ijklmn} - 6\delta^{\lbrack r}_{\lbrack i}\delta^s_j
\delta^{t\rbrack}_{k\rbrack}\eta_{lr}\eta_{ms}\eta_{nt}
\right)
$}

\thicklines
\multiput(20,120)(8,0){3}{\line(1,0){4}}
\put(44,120){\line(1,1){20}}
\put(44,120){\line(1,-1){20}}

\put(64,140){$\Phi^I$}
\put(64,90){$\Phi^I$}
\put(2,115){$H_{ijk}$}
\put(45,135){$p_2$}
\put(45,103){$p_1$}

\thinlines
\put(90,122){\line(1,0){10}}
\put(90,118){\line(1,0){10}}

\put(120,117){$
\Th^{l\lbrack ij}  (p_2)^{k\rbrack} (p_1)_l
+\Th^{l\lbrack ij}  (p_1)^{k\rbrack} (p_2)_l
$}
\end{picture}}

The propagator is:
\bear
\langle H_{ijk}(p) H_{lmn}(-p')\rangle
 &=&
{{36\pi}\over {p^2 + i\epsilon}} 
  \eta_{rl}\eta_{sm}\eta_{tn}
  \left(p_{\lbrack i} p^{\lbrack r} 
    \delta_j^{s}\delta_{k\rbrack}^{t\rbrack}
  -\frac{1}{6}  p_u{\epsilon_{ijk}}^{u\lbrack st} p^{r\rbrack}
\right)
   \delta^{(6)}(p-p')
\nn\\ &&
+{{\pi}\over {2}}\left(
\epsilon_{ijklmn} - 6\delta^{\lbrack r}_{\lbrack i}\delta^s_j
\delta^{t\rbrack}_{k\rbrack}\eta_{lr}\eta_{ms}\eta_{nt}
\right) \delta^{(6)}(p-p').
\nn
\eear

The cubic vertex is:
\bear
\Lambda^{ijk}_{I_1 I_2}(p_1, p_2) &=&
\frac{1}{96\pi}\delta_{I_1 I_2}\Th^{l\lbrack ij}  (p_2)^{k\rbrack} (p_1)_l
+\frac{1}{96\pi}\delta_{I_1 I_2}\Th^{l\lbrack ij}  (p_1)^{k\rbrack} (p_2)_l
\nn
\eear
In terms of $p\equiv p_1 + p_2$ and $q\equiv p_1-p_2$, we can write this
as:
\bear
\Lambda^{ijk}_{I_1 I_2}(p, q) &=&
\frac{1}{192\pi}\delta_{I_1 I_2}\Th^{l\lbrack ij}  p^{k\rbrack} p_l
+\frac{1}{192\pi}\delta_{I_1 I_2}\Th^{l\lbrack ij} q^{k\rbrack} q_l
\nn
\eear

The quartic interaction is:
\bear
A_q &=&
\eta_{pq}\Th^{klp}\Th^{ijq} 
p_{1k} p_{2i} p_{3l} p_{4j}
  (\delta_{I_1 I_4}\delta_{I_2 I_3} -\delta_{I_1 I_2}\delta_{I_3 I_4})
\nn\\ &&
+\eta_{pq}\Th^{klp}\Th^{ijq} 
 p_{1k} p_{2i} p_{3j} p_{4l}
  (\delta_{I_1 I_3}\delta_{I_2 I_4} -\delta_{I_1 I_2}\delta_{I_3 I_4})
\nn\\ &&
+\eta_{pq}\Th^{klp}\Th^{ijq} 
  p_{1k} p_{2l} p_{3i} p_{4j}
  (\delta_{I_1 I_4}\delta_{I_2 I_3} -\delta_{I_1 I_3}\delta_{I_2 I_4})
\nn
\eear

Let us define:
$$
p\equiv p_1 +p_2,\qquad q\equiv p_1-p_2,\qquad
r\equiv p_3-p_4.
$$
We have:
$$
p^2 = -q^2 = -r^2,\qquad
p\cdot q = p\cdot r = 0.
$$
We get $A_s = A'_s \delta_{I_1 I_2}\delta_{I_3 I_4}$ with
\bear
A'_s &=& 
+\frac{\pi}{2}\eta_{cf}\Th^{abc}\Th^{def}p_a r_b p_d r_e
+\frac{\pi}{4}\eta_{cf}\Th^{abc}\Th^{def}p_a q_b p_d q_e
\nn\\ &&
+\frac{\pi}{4}p^2\eta_{be}\eta_{cf}\Th^{abc}\Th^{def} p_a p_d
-\frac{\pi}{16}r_l q^l\eta_{be}\eta_{cf}\Th^{abc}\Th^{def} q_a r_d
\nn\\ &&
+\frac{\pi}{2p^2}\Th^{abc}\Th^{def} p_a q_b r_c p_d q_e r_f
+\frac{\pi}{2p^2}q_l r^l\eta_{cf}\Th^{abc}\Th^{def} p_a q_b p_d r_e.
\nn\\ &&
\label{schan}
\eear

The t-channel is given by the same expression (\ref{schan}) with
$$
p = p_3-p_1,\qquad q = p_3 +p_1,\qquad
r = p_4 + p_2,
$$
and the u-channel is given by (\ref{schan}) with:
$$
p = p_4-p_1,\qquad q = p_4 +p_1,\qquad
r = p_3 + p_2.
$$

\subsection{Large impact parameters}
The approximation of large impact parameter corresponds to
$t\gg s,u$. In this case the $t$-channel amplitude dominates.
We can Fourier transform to obtain a force that behaves, at large
distances, like $\frac{v^2}{r^4}$. 
The $H$-propagator indeed
generates a force that behaves as ${1\over {r^4}}$
(because it is a harmonic function only in the transverse directions).

In this approximation with keep only the t-channel and we set:
\bear
p_1 &=& (\vec{v}-\frac{1}{2}\vec{p},\,
  \frac{R_{\|}}{2}(\vec{v}-\frac{1}{2}\vec{p})^2,\,
          \frac{1}{R_\|}),\nn\\
p_2 &=& (-\vec{v}+\frac{1}{2}\vec{p},\,
  \frac{R_{\|}}{2}(\vec{v}-\frac{1}{2}\vec{p})^2,\,
          \frac{1}{R_\|}),\nn\\
p_3 &=& (\vec{v}+\frac{1}{2}\vec{p},\,
    \frac{R_{\|}}{2}(\vec{v}+\frac{1}{2}\vec{p})^2,\, \frac{1}{R_\|}),\nn\\
p_4 &=& (-\vec{v}-\frac{1}{2}\vec{p},\,
    \frac{R_{\|}}{2}(\vec{v}+\frac{1}{2}\vec{p})^2,\, \frac{1}{R_\|}).\nn
\eear
The notation is:
$$
p_i = (\vec{p}_i,p_{i-},p_{i+}),\qquad
p_i^2 = \vec{p}_i^2 -2 p_{i-} p_{i+}.
$$
and $R_{\|}$ is the radius of the light-like direction of 
  M(atrix)-theory.  We set:
\bear
p &=& (\vec{p},\, R_{\|}(\vec{v}\cdot\vec{p}),\, 0),\nn\\
q &=& (2\vec{v},\, R_{\|}(\vec{v}^2+\frac{1}{4}\vec{p}^2),\,
     \frac{2}{R_\|}),\nn\\
r &=& (-2\vec{v},\, R_{\|}(\vec{v}^2+\frac{1}{4}\vec{p}^2),\,
     \frac{2}{R_\|}),\nn
\eear
and assume that $|\vec{p}|\ll |\vec{v}|$.

Following \cite{ABS,NekSch,Berk}, we set the nonzero components of $\Th$
to be $\Th^{ab+} \equiv \th^{ab}$, with $a,b=1\dots 4$ and
$\th$ is anti-self-dual on $\MR{4}$.
We find that the amplitude is proportional to:
\bear
A &\propto&
\frac{32\pi}{R_{\|}^2 p^2}\th^{ab}p_a v_b \th^{de} p_d v_e
+\frac{6\pi}{R_{\|}^2 p^2}\vec{v}^2\eta_{cf}\th^{ac}\th^{df} p_a p_d.
\eear
After a Fourier transform with respect to $\vec{p}$, this indeed
produces a force that is proportional to $v^2/r^4$.



\def\np#1#2#3{{\it Nucl.\ Phys.} {\bf B#1} (#2) #3}
\def\pl#1#2#3{{\it Phys.\ Lett.} {\bf B#1} (#2) #3}
\def\physrev#1#2#3{{\it Phys.\ Rev.\ Lett.} {\bf #1} (#2) #3}
\def\prd#1#2#3{{\it Phys.\ Rev.} {\bf D#1} (#2) #3}
\def\ap#1#2#3{{\it Ann.\ Phys.} {\bf #1} (#2) #3}
\def\ppt#1#2#3{{\it Phys.\ Rep.} {\bf #1} (#2) #3}
\def\rmp#1#2#3{{\it Rev.\ Mod.\ Phys.} {\bf #1} (#2) #3}
\def\cmp#1#2#3{{\it Comm.\ Math.\ Phys.} {\bf #1} (#2) #3}
\def\mpla#1#2#3{{\it Mod.\ Phys.\ Lett.} {\bf #1} (#2) #3}
\def\jhep#1#2#3{{\it JHEP.} {\bf #1} (#2) #3}
\def\atmp#1#2#3{{\it Adv.\ Theor.\ Math.\ Phys.} {\bf #1} (#2) #3}
\def\jgp#1#2#3{{\it J.\ Geom.\ Phys.} {\bf #1} (#2) #3}
\def\cqg#1#2#3{{\it Class.\ Quant.\ Grav.} {\bf #1} (#2) #3}
\def\hepth#1{{\it hep-th/{#1}}}


\begin{thebibliography}{99}

\bibitem{ABS}{O. Aharony, M. Berkooz and N. Seiberg,
  {``Light-Cone Description of $(2,0)$
   Superconformal Theories in Six Dimensions,''}
   \atmp{2}{1998}{119}, \hepth{9712117}}

\bibitem{NekSch}{N.Nekrasov, A.Schwarz,
  {``Instantons on noncommutative $R^4$ and $(2,0)$
  superconformal six dimensional theory,''} \hepth{9802068}}

\bibitem{Berk}{M. Berkooz,
  {``Non-local Field Theories and the Non-commutative Torus,''}
  \hepth{9802069}}

\bibitem{CDS}{A.~Connes, M.R.~Douglas and A.~Schwarz,
  {``Noncommutative Geometry and Matrix Theory:
  Compactification on Tori,''} \jhep{02}{1998}{003}, \hepth{9711162}}

\bibitem{DH}{M.R.~Douglas and C.~Hull,
  {``D-branes and the Noncommutative Torus,''}
  \jhep{02}{1998}{008}, \hepth{9711165}} 

\bibitem{SWNCG}{N. Seiberg and E. Witten,
  {``String Theory and Noncommutative Geometry,''}
  \jhep{9909}{1999}{032}, \hepth{9908142}}

\bibitem{GMSS}{R. Gopakumar, S. Minwalla, N. Seiberg and A. Strominger,
  {``OM Theory in Diverse Dimensions,''}
  \hepth{0006062}}

\bibitem{HVerl}{H. Verlinde,
  {``A Matrix String Interpretation of the Large N Loop Equation,''}
  \hepth{9705029}}
\bibitem{GuKlPo}{S.~Gukov, I.~R.~Klebanov and A.~M.~Polyakov,
    {``Dynamics of $(n,1)$ strings,''}
    \pl{423}{1998}{64}, \hepth{9711112}}


\bibitem{SSTi}{N. Seiberg, L. Susskind and N. Toumbas,
  {``Space/time non-commutativity and causality,''} \hepth{0005015}}
\bibitem{BBSS}{E. Bergshoeff, D. S. Berman,
  J. P. van der Schaar, P. Sundell
  {``A Noncommutative M-Theory Five-brane,''}
  \hepth{0005026}}
\bibitem{SSTii}{N. Seiberg, L. Susskind and N. Toumbas,
    {``Strings in background electric field,
    space/time noncommutativity and a new noncritical string theory,''}
    \jhep{0006}{2000}{021}, \hepth{0005040}}
\bibitem{GGS}{O.J. Ganor, Govindan Rajesh and S. Sethi,
    {``Duality and noncommutative gauge theories,''} \hepth{0005046}}
\bibitem{GMMS}{R.~Gopakumar, J.~Maldacena, S.~Minwalla and A.~Strominger,
  {``S-duality and noncommutative gauge theory,''}
  \hepth{0005048}} 
\bibitem{BarRab}{J.L.F. Barb\'on and E. Rabinovici,
  {``Stringy fuzziness as the
  custodian of time-space noncommutativity,''} \hepth{0005073}}
\bibitem{GomMeh}{J. Gomis and T. Mehen,
  {``Space-Time Noncommutative Field Theories And Unitarity,''}
  \hepth{0005129}}
\bibitem{ChWu}{G.-S. Chen and Y.-S. Wu,
  {``Comments on Noncommutative Open String Theory: V-duality and
  Holography,''}, \hepth{0006013}}
\bibitem{Harm}{T. Harmark,
  {``Supergravity and Space-Time Non-Commutative Open String Theory,''}
 \hepth{0006023}}
\bibitem{KleMal}{I.R. Klebanov and J. Maldacena,
  {``$1+1$ dimensional NCOS and its $U(N)$ gauge theory dual,''}
  \hepth{0006085}} 
\bibitem{RusShe}{J.J Russo and M.M. Sheikh-Jabbari,
  {``On Noncommutative Open String Theories,''}
  \hepth{0006202}}
\bibitem{AhGoMe}{O. Aharony J. Gomis and T. Mehen,
  {``On Theories With Light-Like Noncommutativity,''}
  \hepth{0006236}}

\bibitem{BFSS}{T. Banks, W. Fischler, S.H. Shenker, L. Susskind,
  {``M Theory As A Matrix Model: A Conjecture,''}
  \prd{55}{1997}{5112}, \hepth{9610043}}

\bibitem{DKPS}{M.R. Douglas, D. Kabat, P. Pouliot and S. Shenker,
  ``D-branes and Short Distances in String Theory,''
  \np{485}{1997}{85}, \hepth{9608024}}

\bibitem{PSS}{S. Paban, S. Sethi and M. Stern,
  {``Constraints From Extended Supersymmetry in Quantum Mechanics,''}
  \np{534}{1998}{137}, \hepth{9805018}}

\bibitem{RamTay}{W. Taylor and M. van Raamsdonk,
  ``Multiple D0-branes in Weakly curved backgrounds,''
  \np{558}{1999}{63},\hepth{9904095}}

\bibitem{GanMot}{O.J. Ganor and L. Motl,
  {``Equations of the (2,0) Theory
  and Knitted Five-Branes,''}
  \jhep{05}{1998}{009}, \hepth{9803108}} 

\bibitem{KZ}{M. Kreuzer, J.-G. Zhou,
  {``$\Lambda$-symmetry and background independence
  of noncommutative gauge theory on $R^n$,''}
    \jhep{0001}{2000}{011}, \hepth{9912174}}

\bibitem{HKLR}{C.M. Hull, A. Karlhede, U. Lindstr\"om and M. Rocek,
  ``Nonlinear $\sigma$-models and their gauging in and out
  of superspace,'' \np{266}{1986}{1}}
  
\bibitem {N=4}{E.E.Donets, A.Pashnev, J.J.Rosales, M.M.Tsulaia,
  {``N=4 Supersymmetric Multidimensional Quantum Mechanics,
  Partial SUSY Breaking and Superconformal Quantum Mechanics,''}
  \prd{61}{2000}{043512}, \hepth{9907224}}

\bibitem {wb}{J.Wess, J.Bagger,
  {\it Supersymmetry and Supergravity}, Princeton University Press, 1991.}

\bibitem {casalbuoni} {R.Casalbuoni, {``The Classical
  Mechanics for Bose-Fermi Systems,''} Nuevo Cimento {\bf A33} (1976) 389}

\bibitem{WIP}{Work in progress.}

\end{thebibliography}
\end{document}